\documentclass[
 reprint,
superscriptaddress,
 amsmath,amssymb,
 aps, prx
]{revtex4-2}

\usepackage{pifont}
\usepackage{graphicx}% Include figure files
\usepackage{bm}% bold math
\usepackage{sidecap}
\usepackage{xcolor}
\usepackage[colorlinks=true,citecolor=blue]{hyperref}
\usepackage{natbib}

\makeatletter

\newcommand{\Rmnum}[1]{\expandafter\@slowromancap\romannumeral #1@}
\makeatother

\begin{document}

\title{Lanthanum Oxyhalide Monolayers: \\ An Exceptional Dielectric Companion to Two-Dimensional Semiconductors}% Force line breaks 

\author{Zhuoling Jiang}
\affiliation{Science, Mathematics and Technology (SMT), Singapore University of Technology and Design, Singapore 497372}

\author{Tong Su}
\affiliation{Science, Mathematics and Technology (SMT), Singapore University of Technology and Design, Singapore 497372}

\author{Cherq Chua}
\affiliation{Science, Mathematics and Technology (SMT), Singapore University of Technology and Design, Singapore 497372}

\author{L. K. Ang}
\affiliation{Science, Mathematics and Technology (SMT), Singapore University of Technology and Design, Singapore 497372}

\author{Chun Zhang}
\affiliation{Department of Physics, National University of Singapore, Singapore 117542}

\author{Liemao Cao}
\affiliation{College of Physics and Electronic Engineering, Hengyang Normal University, Hengyang 421002, China}

\author{Yee Sin Ang}
\email{yeesin\_ang@sutd.edu.sg}
\affiliation{Science, Mathematics and Technology (SMT), Singapore University of Technology and Design, Singapore 497372}

\begin{abstract}
Two-dimensional (2D) layered dielectrics offers a compelling route towards the design of next-generation ultimately compact nanoelectronics. Motivated by recent high-throughput computational prediction of LaO$X$ ($X$ = Br, Cl) as an exceptional 2D dielectrics that significantly outperforms HfO$_2$ even in the monolyaer limit, we investigate the interface properties between LaOX and the archetypal 2D semiconductors of monolayer transition metal dichacolgenides (TMDCs) $M$S$_2$ ($M$ = Mo, W) using first-principle density functional theory simulations. We show that LaO$X$ monolayers interacts weakly with $M$S$_2$ via van der Waals forces with negligible hybridization and interfacial charge transfer, thus conveniently preserving the electronic properties of 2D TMDCs upon contact formation. The conduction and valance band offsets of the interfaces exhibit a sizable value ranging from 0.7 to 1.4 eV, suggesting the capability of LaO$X$ as a gate dielectric materials. Based on Murphy-Good electron emission model, we demonstrate that LaOCl/MoS$_2$ is a versatile dielectric/semiconductor combinations that are compatible to both NMOS and PMOS applications with leakage current lower than $10^{-7}$ Acm$^{-2}$, while LaO$X$/WS$_2$ is generally compatible with PMOS application. The presence of an interfacial tunneling potential barrier at the van der Waals gap further provide an additional mechanism to suppress the leakage current. Our findings reveal the role LaO$X$ as an excellent dielectric companion to 2D TMDC and shall provide useful insights for leveraging the dielectric strength of LaO$X$ in the design of high-performance 2D nanodevices. 

\end{abstract}

\maketitle

\section{Introduction}
Two-dimensional materials represent an emerging platform for the exploration of next-generation solid-state device technology beyond the `silicon era' \cite{2D_si}. 
The ever-expanding 2D family, with over 4000 species predicted to date \cite{database1,database2}, covers almost the entire spectrum of technologically important solid-states phases, including dielectric, semiconductor, semimetal, metal and superconductors, as well as exotic condensed matter phases, such as topological insulator (TI), Dirac semimetal \cite{semimetal}, Mott insulator \cite{mott}, hourglass semimetal \cite{hourglass} and so on.
Myriads of novel functional devices, such as computing electronics \cite{circuit, adv_rev}, neuromorphic devices \cite{neuro}, energy-efficient electronics \cite{AEM_FET}, reversible logical cells \cite{valley}, terahertz devices \cite{graphene_NOR1, graphene_NOR2}, renewable energy converters \cite{wang_rev} and photodetectors \cite{avalanche}, have been concretely demonstrated in the past decade. 

2D layered dielectric materials have also been actively explored recently for their potential in improving the reliability of field-effect transistor (FET) \cite{2D_rev1}. 
Hexagonal boron nitride (hBN) exhibits exceptionally large breakdown electric field reaching 21 MV cm$^{-1}$ at a thickness of 3 nm \cite{hBN_break} and its 2D layered morphology has led to unusual features in charge transport \cite{hBN_SCLC1, hBN_SCLC2, hBN_tunnel}, dielectric breakdown mechanism \cite{hBN_layer} and random telegraphic noises when compared to bulk oxides \cite{hBN_tele}. 
However, the relatively low out-of-plane static dielectric constant of $\epsilon_{\perp,0} = 3.29$ in hBN has severely limited their potential as an efficient dielectric materials \cite{hBN_dielectric}. 
Beyond hBN, recent high-throughput calculation has uncovered 32 species of 2D layered dielectric materials that are stable, non-soluble solid, nontoxic, noncorrosive, and with sufficiently wide band gap for gate dielectric applications \cite{NC2021}. 
Remarkably, the lanthanum oxyhalide monolayers LaO$X$ ($X$=Cl, Br) has been predicted to outperform HfO$_2$ even in the monolayer limit of sub-1-nm thickness with exceedingly high out-of-plane static dielectric constant (55.8 and 13.2, respectively, for LaOCl and LaOBr) and low equivalent oxide thickness (EOT) (0.05 nm and 0.23 nm, respectively, for LaOCl and LaOBr).
The leakage currents, inclusive of both field-induced tunneling and overbarrier thermionic emission, are predicted to be as low as $\sim 10^{-7}$ (LaOCl) and $\sim10^{-11}$ Acm$^{-2}$ for a specific hypothetical semiconductor of conduction band minimum (CBM) and valance band maximum (VBM) energies of -2 eV and -4 eV, respectively. 
Such ultralow leakage current significantly outperforms the requirement of $\mathcal{J}_{\text{leakage}} < 10^{-2}$ Acm$^{-2}$ for high-performance device applications \cite{2D_rev2}.
The high dielectric constant, small EOT, and the low leakage current suggests that LaO$X$ is a compelling 2D layered dielectric materials for transistor applications. 

\begin{figure*}
\centering
\includegraphics[width=16.58 cm]{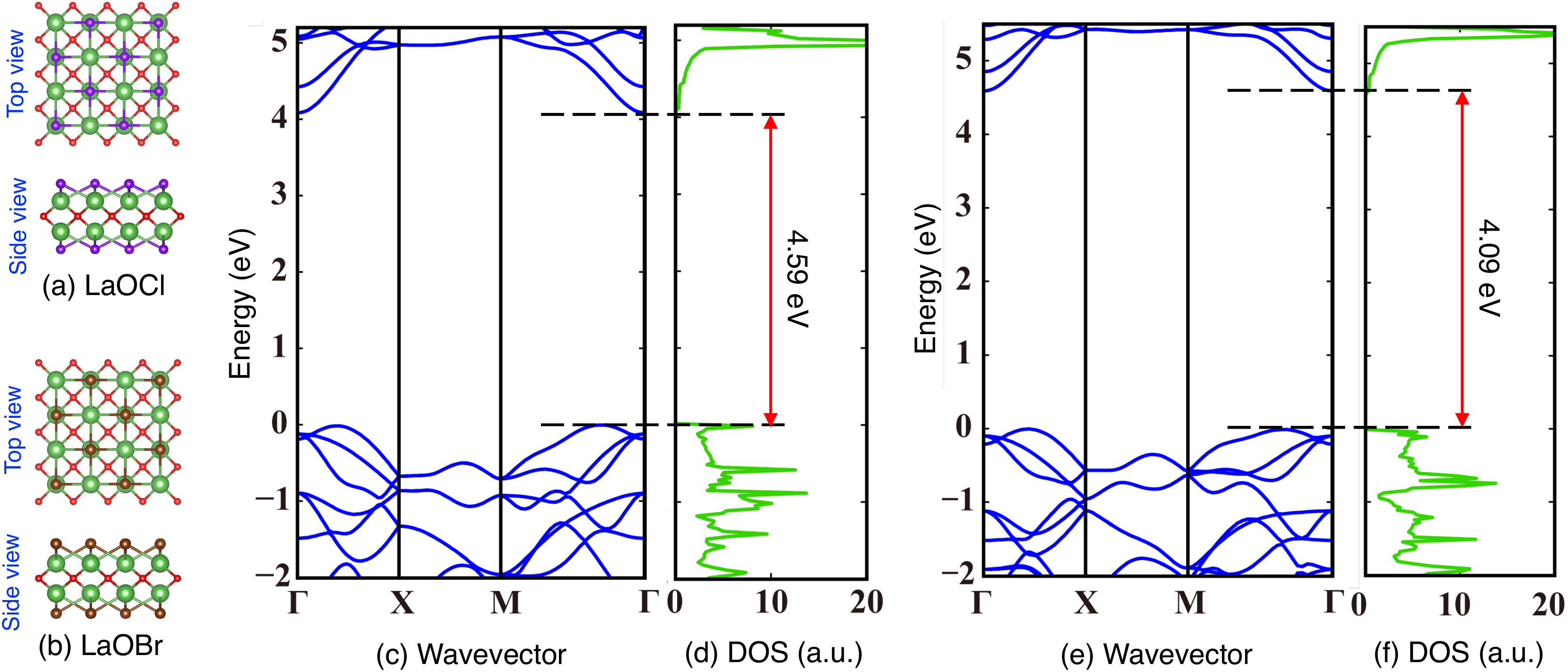}
\caption{\label{fig:1}  Lattice and electronic structures of LaOCl and LaOBr monolayers. (a) and (b) shows the side and top views of the lattice structures of LaOCl and LaOBr, respectively. (c) and (d) shows the electronic band structure and electronic density of states of LaOCl calculated at the PBE level. (e) and (f), same as (c) and (d), but for LaOBr. }
\label{fig:1}
\end{figure*}

Despite enormous recent efforts devoted to understand the interfacial physics of metal/semiconductor and semiconductor/semiconductor van der Waals heterostructures (vdWHs) in which the constituent monolayers are coupled together via weak van der Waals forces \cite{vdW_rev}, 2D/2D dielectric/semiconductor heterostructures remains largely unexplored thus far \cite{2D_vdw_dielectrics, 2D_vdw_dielectric2}. 
Importantly, the interlayer interactions in vdWH, such as orbital hybridization and interfacial charge transfer, can substantially modify the electronic band structures and transport properties of the semiconducting channel, and can uncontrollably alter the band alignment of the dielectric/semiconductor interface, thus complicating and/or compromising the design and performance of 2D semiconductor devices.
In relevance to LaO$X$ monolayers, although their \textit{free-standing} monolayer dielectric properties have been comprehensively catalogued \cite{NC2021}, their \textit{interfacial physics and overall compatibility with 2D semiconductors as a gate dielectric} remains an open question.
Particularly, the electronic structures and band offsets of LaO$X$/2D-semiconductor contact heterostructures are critical quantities that needs to be characterized before the potential usefulness of LaO$X$ in 2D semiconductor device technology can be fully appreciated. 

In this work, we perform first-principle density functional theory (DFT) simulations of 2D/2D dielectric/semiconductor vdWHs composed of LaO$X$ monolayer in contact with the \textit{archetypal} 2D semiconductors, transition metal dichalcogenide (TMDC) monolayers MoS$_2$ and WS$_2$, $M$S$_2$ ($M$ = Mo, W). 
Four contact heterostructures, i.e. LaOCl/MoS$_2$, LaOCl/WS$_2$, LaOBr/MoS$_2$ and LaOBr/WS$_2$, are computationally designed and investigated.
Remarkably, we found that LaOC$X$ are exceptional \textit{dielectric companion} to MoS$_2$ and WS$_2$ monolayers via the formation of `clean' interface with extremely weak interlayer interaction and negligible interfacial charge transfer across of $10^{-2}$ $e$ per supercell. 
Such weak interaction enables the electronic states around the conduction band minima (CBM) and valance band maxima (VBM) of 2D TMDC to be completely preserved. 
Importantly, because the CBM and VBM energies of 2D TMDC are conveniently situated around the mid-gap regime of LaOX, both the conduction band offset (CBO) and the valance band offset (VBO) exhibit a sizable magnitude that ranges from 0.7 eV to 1.4 eV. 
Based on Murphy-Good electron emission model \cite{MG}, we demonstrate that LaO$X$/WS$_2$ is generally compatible with PMOS device configuration with leakge current lower than $10^{-2}$ Acm$^{-2}$.
Intriguingly, LaOCl/MoS$_2$ has an exceptionally low leakage current below $10^{-7}$ $Acm^{-2}$, thus suggesting its compatibility with both NMOS and PMOS applications. 
Our findings reveals LaO$X$ as a compelling gate dielectric companion to 2D TMDC for high-performance nanoelectronic device applications.

\section{Computational Details}

All first-principles calculations based on the density functional theory (DFT) were carried out by using the Vienna ab Initio Simlation Package (VASP) \cite{kresse}. The Projector augmented wave pseudopotentials \cite{blochl} for the core and the Perdew-Burke-Ernzerhof (PBE) format \cite{PBE} of the generalized gradient approximation (GGA) for the exchange-correlation functional were adopted. We set the kinetic energy cutoff to be 500 eV, and optimized the structures until the forces were below 0.01 eV/\AA. The LaOX/MS2 (X=Br, Cl; M=Mo,W) heterostructures are consisting of a $2\sqrt{2} \times 2\sqrt{2}$ cell of LaOX and a $\sqrt{13} \times \sqrt{13}$ cell of MS2. To eliminate the interaction between the slab surfaces, the vacuum thickness was set to be 18 \AA. A $5\times5\times1$ k-point grid was sampled in the Brillouin Zone of the supercell. We have employed the Grimme’s scheme (DFT-D3)\cite{grimme} to describe the interlayer van der Waals interaction, and the dipole correction was also considered. Bader charge calculations were performed to analyze the interlayer charge transfer \cite{henkelman}. 

\begin{figure*}
\centering
\includegraphics[width=15.58cm]{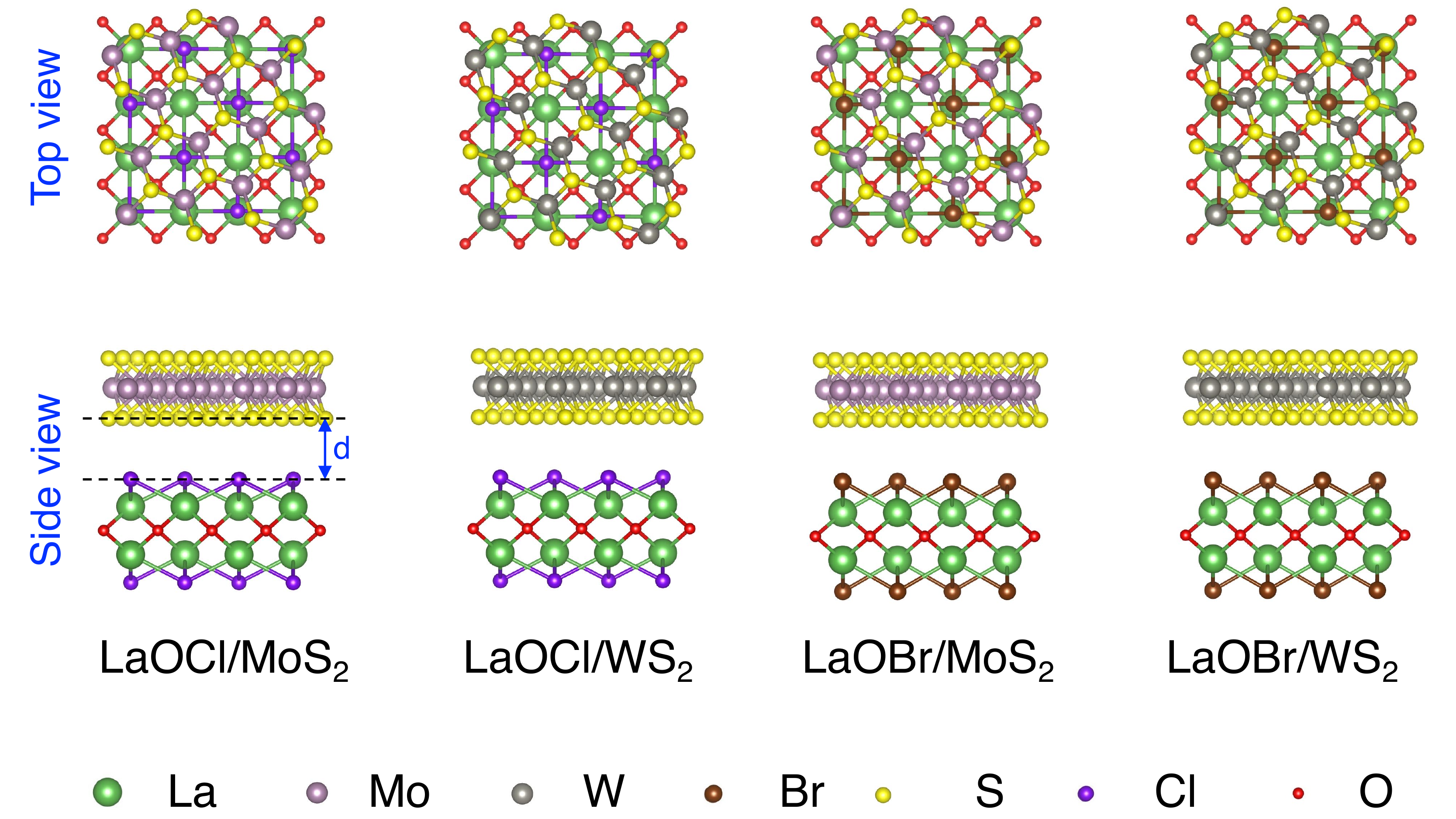}
\caption{\label{fig:contact-geometry}  Lattice structures and stacking configurations of LaO$X$/$M$S$_2$ van der Waals heterostructures. The heterostructures are consisting of $2\sqrt \times 2\sqrt{2}$ cell of LaO$X$ monolayer and $\sqrt{13} \times \sqrt{13}$ cell of $M$S$_2$. All heterostructures supercell contain 93 atoms. }
\label{fig:2}
\end{figure*}

\section{Results and Discussions}

\subsection{Structural and Electronic Properties of LaOCl and LaOBr Monolayers}

The lattice structure and the electronic band structures of LaO$X$ are shown in Fig. 1.
Structurally, LaOX is a tetragonal lattice with a central oxygen layer sandwiched by La atoms. The La atoms are covered by an outer layer composed of halogen atoms [see Figs. 1(a) and (b)]. 
The lattice constants of LaOCl and LaOBr are $3.99$ \AA and $4.01$ \AA, respectively. 
The electronic band structure and density of states (DOS) of freestanding LaOCl [Figs. 1(c) and 1(d)] and LaOBr [Figs. 1(e) and 1(f)] suggests that an indirect band gap of 4.59 eV and 4.09 eV, respectively, which is consistent with the values reported in previous work \cite{NC2021} (see Table I for summary of key parameters of the structural and electronic properties of monolayer LaOCl, LaOBr, MoS$_2$ and WS$_2$). 
We note that as the band gap values are larger than 4.0 eV at the PBE level, LaOCl and LaOBr can be categorized as dielectric materials. 
Particularly for gate dielectric applications, previous machine-learning-based investigation has uncovered a semi-empirical expression for the intrinsic dielectric breakdown field strength \cite{ML1}, $\mathcal{E}_b = 24.442 \exp\left(0.315 \sqrt{E_g \omega_{max}}\right)$, where $E_g$ is the band gap of the dielectric material, and $\omega_{max}$ is the maximum phonon frequency. 
The large band gap values of LaOCL and LaOBr are thus beneficial for achieving high dielectric breakdown strength.
We further note that as the constructed 2D/2D dielectric/semiconductor vdWH contains large number of atoms (i.e. 93 atoms in the supercell), PBE method, which underestimate the band gap as compared to the more computationally expensive HSE06 hybrid functional method (i.e. 5.83 eV and 5.68 eV, respectively, for LaOCl and LaOBr \cite{NC2021}), is employed in this work due to computational resource constraint. 
Nevertheless, the PBE calculation is expected to yield the same qualitative physics insights for 2D material vdWHs \cite{beyond_AR}. 

\bgroup
\def\arraystretch{1.5}
\begin{table}[]
\centering
\caption{Lattice constant ($a$), CBM energy ($E_{\text{CBM}}$), VBM energy ($E_{\text{VBM}}$), and band gap ($E_g$) of the LaO$X$ and $M$S$_2$ monolayers.  }
\label{table:ML}
\begin{tabular}{ccccc}
\hline
%\multirow{2}{*}{$l$} & \multicolumn{2}{c}{$\eta$ $(L/D=50)$} & \multicolumn{2}{c}{$\eta$ $(L/D=100)$} \\ \cline{2-5} 
Monolayer          &  $a$ (\AA)  & $E_{\text{CBM}}$ (eV) & $E_{\text{VBM}}$ (eV) & $E_g$ (eV)\\ \hline
LaOCl               & 3.99        & -2.07        & -6.66      & 4.59      \\
LaOBr               & 4.04        & -2.46        & -6.65      & 4.09      \\
MoS$_2$             & 3.18        & -4.23        & -5.93      & 1.70      \\
WS$_2$              & 3.18        & -3.84        & -5.68      & 1.84      \\ \hline
\end{tabular}
\end{table}
\egroup

\begin{figure*}
\centering
\includegraphics[scale=0.8058]{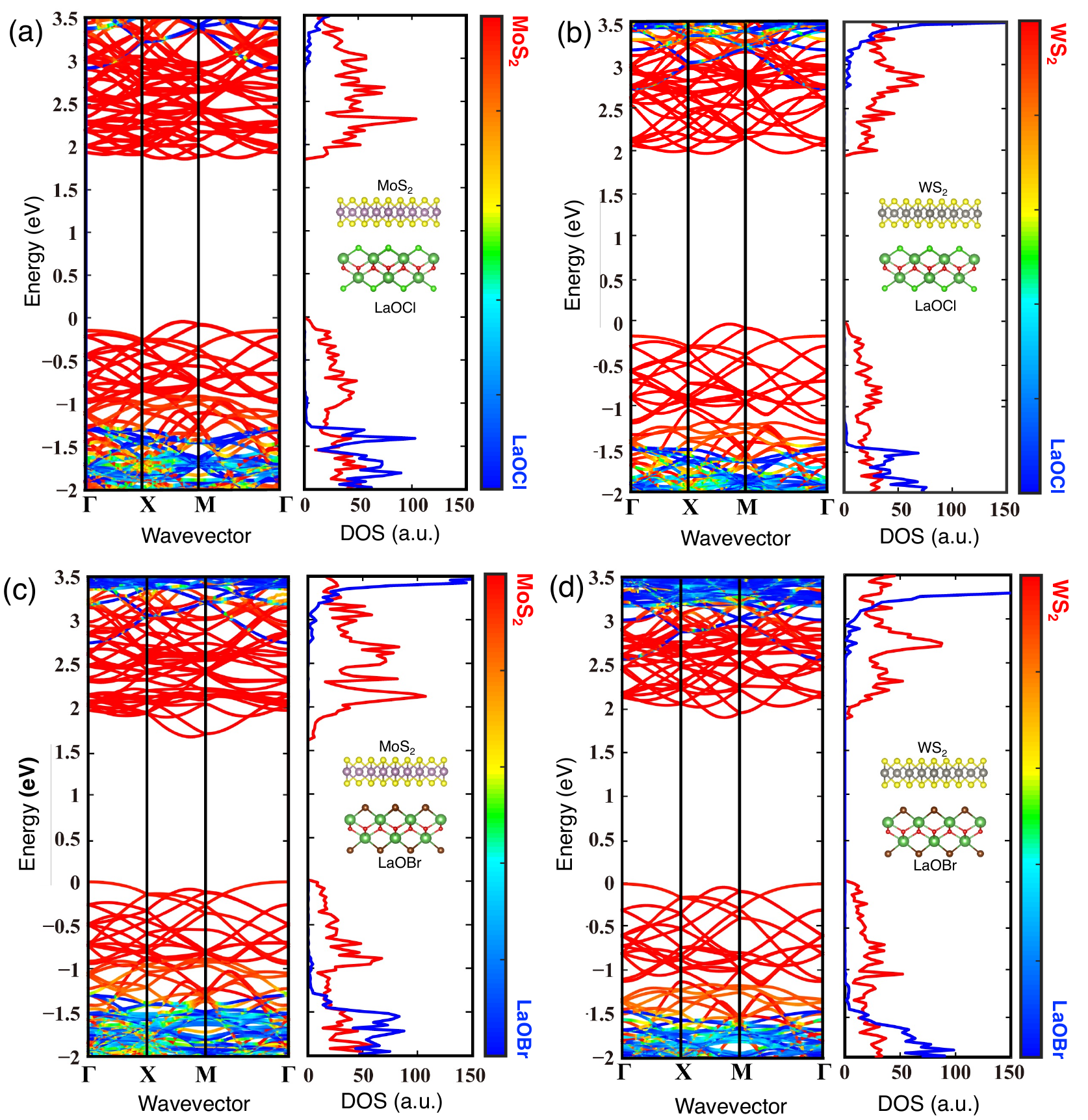}
\caption{\label{fig:3}  Electronic band structures (left panel) of LaO$X$/$M$S$_2$ for (a) LaOCl/MoS$_2$; (b) LaOCl/WS$_2$; (c) LaOBr/MoS$_2$; and (d) LaOBr/WS$_2$. The right panel shows the density of states projected to LaO$X$ and $M$S$_2$ monolayers. }
\label{fig:3}
\end{figure*}

\bgroup
\def\arraystretch{1.5}
\begin{table*}[t]
\centering
\caption{Lattice mismatch ($\epsilon$), binding energy ($E_B$), interlayer distance ($d$), CBM energy ($E_{\text{CBM}}$), VBM energy ($E_{\text{VBM}}$), band gap ($E_g$), CBO ($E_{\text{CBO}}^{(\text{AR})}$) and VBO ($E_{\text{VBO}}^{(\text{AR})}$) calculated via Anderson rule, as well as the CBO ($E_{\text{CBO}}^{(\text{DFT})}$) and VBO ($E_{\text{VBO}}^{(\text{DFT})}$) of LaO$X$/$M$S$_2$ vdWHs.  }
\label{table:2}
\begin{tabular}{ccccccccccc}
\hline
%\multirow{2}{*}{$l$} & \multicolumn{2}{c}{$\eta$ $(L/D=50)$} & \multicolumn{2}{c}{$\eta$ $(L/D=100)$} \\ \cline{2-5} 
vdWH &  $\epsilon$ (\%) & $E_b$ (eV) & $d$ (\AA) & $E_{\text{CBM}}$ (eV) & $E_{\text{VBM}}$ (eV) & $E_g$ (eV) & $E_{\text{CBO}}^{(\text{AR})}$ & $E_{\text{VBO}}^{(\text{AR})}$ & $E_{\text{CBO}}^{(\text{DFT})}$ & $E_{\text{VBO}}^{(\text{DFT})}$\\ \hline
LaOCl/MoS$_2$ & 1.66 & -2.35 & 3.14 & -3.97 & -5.88 & 1.91 & 2.16 & 0.73 & 1.21 & 1.10\\
LaOCl/WS$_2$ & 1.66 & -2.49 & 3.18 & -3.68 & -5.69 & 2.01 & 1.77 & 0.98 & 0.80 & 1.41\\
LaOBr/MoS$_2$ & 0.78 & -2.38 & 3.21 & -4.18 & -5.85 & 1.67& 1.77 & 0.62 & 1.10 & 0.85\\
LaOBrWS$_2$  & 0.78 & -2.38 & 3.26 & -3.76 & -5.68 & 1.92 & 1.38 & 0.87 & 0.74 & 1.15\\ \hline
\end{tabular}
\end{table*}
\egroup

For both LaOCl and LaOBR, the CBM is situated at the $\Gamma$ point while the VBM is situated along the $\Gamma$-$M$ and the $\Gamma$-$X$ directions. 
The CBM and VBM energies are -2.07 eV and -6.66 eV, respectively, for LaOCl and are -2.46 eV and -6.55 eV, respectively, for LaOBr, where vacuum level is defined as zero. 
Remarkably, the CBM and VBM energies of MoS$_2$ (i.e. -4.23 eV and -5.93 eV, respectively) and WS$_2$ (i.e. -3.84 eV and -5.68 eV, respectively) conveniently located around the middle portion of the band gap of both LaOCl and LaOBr, suggesting that sizable CBO and VBO can be achieved. 
Using the Anderson rule \cite{AR1} which provides a simple estimation of the band offsets by aligning the vacuum level of two contacting materials \cite{AR_TMDC}, the magnitude of the CBO and VBO energies can be estimated as 
\begin{subequations}
    \begin{equation}
    E_{\text{CBO}}^{\text{(AR)}} = \left|E_{\text{CBM},1}^{\text{(ML)}} - E_{\text{CBM},2}^{\text{(ML)}} \right|,
\end{equation}
\begin{equation}
    E_{\text{VBO}}^{\text{(AR)}} = \left|E_{\text{VBM},1}^{\text{(ML)}} - E_{\text{VBM},2}^{\text{(ML)}} \right|,
\end{equation}
\end{subequations}
respectively, where 
$E_{\text{CBM},i}^{\text{(ML)}}$ and $E_{\text{VBM}, i}^{\text{(ML)}}$ are the CBM and VBM energies of the $i$th monolayer ($i=1,2$ for bilayer vdWH), respectively. 
Accordingly, the band offsets ranges from the lowest value of $E_{\text{VBO}}^{\text{(AR)}} = 0.62$ eV for LaOBr/MoS$_2$ to the highest value of $E_{\text{CBO}}^{\text{(AR)}} = 2.16$ eV for LaOCl/MoS$_2$. 
The sizable CBO and VBO estimated via Anderson rule provide a preliminary assurance on the feasibility of LaOCl and LaOBr as gate dielectrics to MoS$_2$ and WS$_2$ monolayers, which will be further verified via direct DFT calculations in the following sections.

\begin{figure*}
\centering
\includegraphics[scale=0.5258]{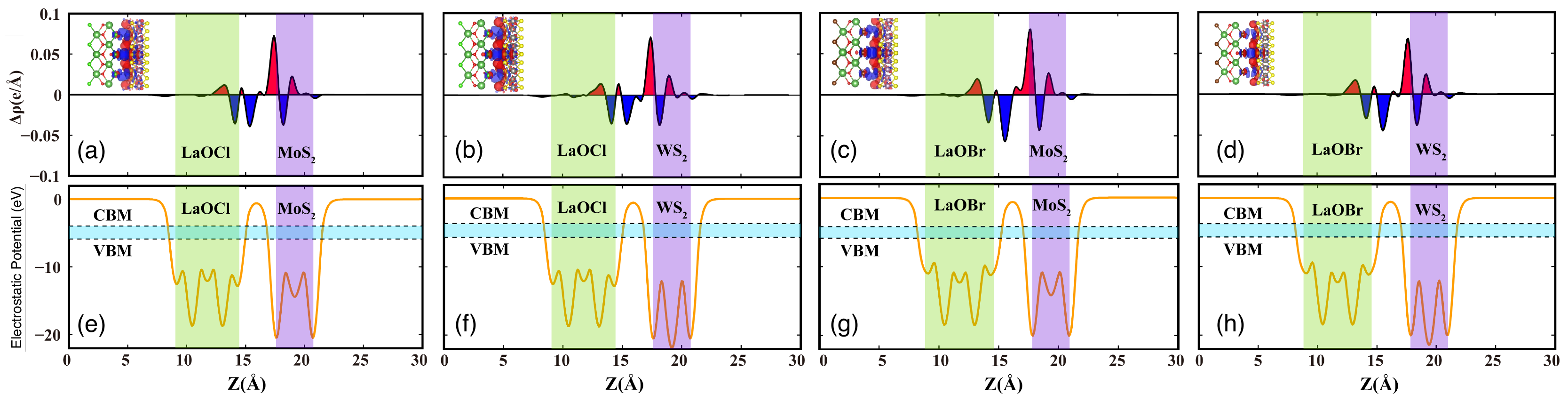}
\caption{\label{fig:4} Interfacial electronic properties of LaO$X$/$M$S$_2$ heterostructures. (a) to (d) shows the plane-averaged differential charge density plots for LaOCl/MoS$_2$, LaOCl/WS$_2$, LaOBr/MoS$_2$ and LaOBr/WS$_2$, respectively. (e) to (h) same as (a) to (d) but for the plane-averaged electrostatic potential profile of the heterostructures. }
\label{fig:4}
\end{figure*}

\bgroup
\def\arraystretch{1.5}
\begin{table*}[t]
\centering
\caption{Interfacial dipole potential ($\Delta V$), interfacial charge transfer ($\Delta q$), TB height relative to CBM ($\Phi_{\text{TB}}^{\text{CBM}}$), TB width relative to CBM ($w_{\text{TB}}^{\text{CBM}}$), TB height relative to VBM ($\Phi_{\text{TB}}^{\text{VBM}}$) and TB width relative to CBM ($w_{\text{TB}}^{\text{VBM}}$) of LaO$X$/$M$S$_2$ vdWHs.  }
\label{table:3}
\begin{tabular}{ccccccc}
\hline
%\multirow{2}{*}{$l$} & \multicolumn{2}{c}{$\eta$ $(L/D=50)$} & \multicolumn{2}{c}{$\eta$ $(L/D=100)$} \\ \cline{2-5} 
vdWH &  $\Delta V$ (eV) & $\Delta q$ (e) & $\Phi_{\text{TB}}^{\text{VBM}}$ (eV) & $w_{\text{TB}}^{\text{CBM}}$ (\AA) & $\Phi_{\text{TB}}^{\text{CBM}}$ (eV) & $w_{\text{TB}}^{\text{CBM}}$ (\AA) \\ \hline
LaOCl/MoS$_2$ & -0.03 & 0.05 & 5.28 & 1.92 & 3.37 & 1.69 \\
LaOCl/WS$_2$ & -0.06 & 0.04 & 5.07 & 1.88 & 3.06 & 1.63 \\
LaOBr/MoS$_2$ & 0.04 & 0.08 & 5.24 & 1.92 & 3.57 & 1.70 \\
LaOBrWS$_2$  & 0.01 & 0.06 & 5.11 & 1.95 & 3.19 & 1.74 \\ \hline
\end{tabular}
\end{table*}
\egroup

\subsection{Structural and Electronic Properties of LaO$X$/$M$S$_2$ Heterostructures}

Four 2D/2D dielectric/semiconductors are constructed: (i) LaOCl/MoS$_2$; (ii) LaOCl/WS$_2$; (iii) LaOBr/MoS$_2$; and (iv) LaOBr/WS$_2$. The top and side views of the fully relaxed structures are shown in Fig. 2 (see Table 2 for a summary of the key parameters calculated for LaO$X$/$M$S$_2$ heterostructures). 
Upon forming the heterostructure, a strain of 1.66\% and 0.78\% is evenly distributed among the two monolayers for LaOCl/$M$S$_2$ and LaOBr/$M$S$_2$ heterostructures, respectively, due to lattice mismatch.  
The interlayer distances of the heterostructures range between 3.1 \AA to 3.4 \AA, thus suggesting that the constituent monolayers are coupled weakly via van der Waals forces. 
The binding energy of the heterostructure can be calculated as \cite{Eb}:
\begin{equation}
    E_b = E_{\text{vdWH}} - E_{\text{LaO}X} - E_{M\text{S}_2}
\end{equation}
where $E_{\text{vdWH}}$, $E_{\text{LaO}X}$ and $E_{M\text{S}_2}$ are the total energy of the combined vdWH, LaO$X$ and $M$S$_2$, respectively. 
As summarized in Table II, all heterostructures have $E_b \approx - 2.4 \sim 2.5$ eV after full structural relaxation, hus confirming their energetic stability. 

In Fig. 3, the electronic band structures and the DOS projected to the constituent monolayers are shown. 
The electronic bands of M$_2$ and LaO$X$ are clearly can be clearly distinguished around the band edges, thus suggesting minimal hybridization of the two materials. 
The band gap of LaOCl/MoS$_2$ and LaOCl/WS$_2$ are 1.91 eV and 2.01 eV, respectively, which are slightly enhanced as compared to that of freestanding MoS$_2$ and WS$_2$ (i.e. 1.70 eV and 1.84 eV, respectively, due to the straining of the 2D semiconductors. 
For LaOBr/MoS$_2$ and LaOBr/WS$_2$ heterostructures, because of the smaller lattice mismatch of $<1$ \%, the band gap values are 1.67 eÍV and 1.92 eV, respectively, which are much closer to the freestanding monolayer band gap values. 
%A particularly important is the existence of a more prominent 'tail states' just above the VBM of LaOCl and LaOBr which can impact the extracted values of the VBOs. 
The CBO and VBO can be extracted from the DOS via 
\begin{subequations}
\begin{equation}
    E_{\text{CBO}}^{\text{(DFT)}} = \left | E_{\text{CBM}, 1}^{(\text{vdWH})} - E_{\text{CBM}, 2}^{(\text{vdWH})}\right |, 
\end{equation} 
\begin{equation}
    E_{\text{VBO}}^{\text{(DFT)}} = \left | E_{\text{VBM}, 1}^{(\text{vdWH})} - E_{\text{VBM}, 2}^{(\text{vdWH})}\right |,
\end{equation}
\end{subequations}
respectively. 
The band offsets obtained from direct DFT calculations ranges from $E_{\text{VBO}}^{\text{(DFT)}} = 0.74$ eV for LaOBr/WS$_2$ to $E_{\text{CBO}}^{\text{(DFT)}} = 1.41$ eV for LaOCl/WS$_2$ (see Table II). 
The band offsets obtained via direct DFT calculations are quantitatively different from those obtained via Anderson rule since DFT calculations explicitly capture the interlayer interactions between the two constituent monolayers \cite{AR2}. 
Furthermore, the presence of strain due to lattice mismatch also modulates the band gap of both LaO$X$ and $M$S$_2$ monolayers. 
Particularly, LaOCl/MoS$_2$ are found to be $>1$ eV in both CBO and VBO. 
As MoS$_2$ is a widely studied 2D semiconductors capable of achieving exceptionally low contact resistance \cite{nat_bi} and strongly suppressed Fermi level pinning effect \cite{nat_SM}, the sizable CBO and VBO of LaOCl/MoS$_2$ reveal LaOCl as an exceptional dielectric companion that can further enhance the device performance and reliability of 2D MoS$_2$ nanoelectronic devices. 

\subsection{Interfacial Charge Transfer and Tunneling-Specific Interface Resistivity}

To further elucidate the interface properties of LaO$X$/$M$S$_2$, we calculate the plane-averaged differential charge density ($\delta \rho$) across the contact heterostrucures [Figs. 4(a) to 4(d)]. 
In general, electron is transferred from the LaO$X$ monolayer to the $M$S$_2$ monolayer due to the significantly lower electron affinity of $M$S$_2$. 
The differential charge density are smaller than 0.1 e/\AA across the heterostructure, suggesting that the interfacial charge transfer is negligible. 
Based on Bader charge analysis \cite{henkelman}, the calcualted charge transfer between the monolayer is less than 0.1 e for all vdWHs (see Table 1). 
The ultralow interfacial charge transfer is also reflected in the plane-averaged electrostatic profile of the contact heterostructures [see Figs. 4(e) to 4(h)]. The interfacial dipole potential ($\Delta V$), which arises due to electron accumulaton on one monolayer and electron depletion on the other monolayer, can be calculated as the vacumm level difference on both sides of the vdWH. 
We found that $\Delta V$ is typically in the order of $\sim 10^{-2}$ eV for all LaO$X$/$M$S$_2$ vdWHs, which is consistent with the overall low ($\delta \rho$) observed in Figs. 4(a) to 4(d).

\bgroup
\def\arraystretch{1.5}
\begin{table*}[t]
\centering
\caption{Gate dielectric performance characteristics of LaO$X$/$M$S$_2$ heterostructures. Tunneling-specific interface resistivity relative to CBM ($\rho_t^{\text{CBM}}$) and to VBM ($\rho_t^{\text{VBM}}$)of LaO$X$/$M$S$_2$ vdWHs, and the leakage current calculated from CBO ($\mathcal{J}_{\text{MG}}^{\text{CBO}}$) and VBO ($\mathcal{J}_{\text{MG}}^{\text{VBO}}$) energies using the Murphy-Good electron emission model, and that calculated using the simplified model ($\mathcal{J}_{0}^{\text{CBO}}$ and $\mathcal{J}_{0}^{\text{VBO}}$). }
\label{table:4}
\begin{tabular}{ccccccc}
\hline
%\multirow{2}{*}{$l$} & \multicolumn{2}{c}{$\eta$ $(L/D=50)$} & \multicolumn{2}{c}{$\eta$ $(L/D=100)$} \\ \cline{2-5} 
vdWH & $\rho_t^{\text{VBM}}$ ($\times 10^{-9} \Omega$cm $^{2}$) & $\rho_t^{\text{VBM}}$ ($\times 10^{-9} \Omega$cm $^{2}$) & $\mathcal{J}_{\text{MG}}^{\text{CBO}}$ (Acm$^{-2}$) & $\mathcal{J}_{\text{MG}}^{\text{VBO}}$ (Acm$^{-2}$) & $\mathcal{J}_{\text{0}}^{\text{CBO}}$ (Acm$^{-2}$) & $\mathcal{J}_{\text{0}}^{\text{VBO}}$ (Acm$^{-2}$) \\ \hline
LaOCl/MoS$_2$ & 1.62 & 0.47 & $2.33\times10^{-7}$ & $4.00\times10^{-8}$ & $3.48\times 10^{-9}$ & $9.26\times 10^{-11}$ \\
LaOCl/WS$_2$ & 1.35 & 0.37 & $6.20\times 10^{-2}$ & $1.34\times10^{-13}$ &$7.18\times 10^{-2}$ & $5.66\times10^{-16}$ \\
LaOBr/MoS$_2$ & 1.60 & 0.51 & $1.05\times10^{-6}$ & $7.50\times 10^{-2}$ & $2.70\times 10^{-9}$ & $1.44\times 10^{-3}$ \\
LaOBrWS$_2$  & 1.67 & 0.50 & $9.3\times10^{-1}$ & $9.62\times 10^{-7}$ & $1.58\times 10^{-2}$ & $2.48\times 10^{-9}$ \\ \hline
\end{tabular}
\end{table*}
\egroup

A tunneling barrier (TB) is commonly formed across the van der Waals gap (vdWG) separating the two monolayers \cite{prx}. The height and width of the TB can be directly estimated from the plane-averaged electrostatic profiles of the vdWH.
In metal/semiconductor contacts, the presence of TB in the vdWG substantially lowers the charge injection efficiency and is often more desirable to eliminate the vdW TB as much as possible. 
However, for gate dielectric applications in dielectric/semiconductor contact, vdW TB provide an additional mechanism, in addition to band offsets, to suppress the leakge current.  
Unlike metal/semiconductor contact in which the TB height and width are extracted based on the intersection of the Fermi level and the electrostatic potential profile, the Fermi level in dielectric/semiconductor contact cannot be straightforwardly defined as $E_F$ is material-and/or device-dependent \cite{prx}. 
Instead, we characterize the height and width by using the intersection of the CBM and VBM with the electrostatic profile [see Figs. 4(e) to 4(h)] since the electronic states around the CBM and VBM are dominantly responsible for carrier conduction in $n$-type and $p$-type device, respectively.
The extracted TB heights and widths can be used to calculate the electron tunneling probability based on the WKB formalism \cite{cao_APL}, 
\begin{equation}
    \mathcal{T} = \exp\left( - \frac{4\pi w_{\text{TB}}\sqrt{2m\Phi_{\text{TB}}}}{h}\right),
\end{equation}
where $m$ is the free electron mass, $h$ is the Planck constant, $w_{\text{TB}}$ and $\Phi_{\text{TB}}$ represents the TB width and height, respectively. 
Here $\mathcal{T}$ is an important quantity characterizing the charge injection efficiency across the vdWG since the ballistic tunneling current and the tunneling-specific interface resistivity across the vdWG are both directly related to $\mathcal{T}$ \cite{cao_APL}.
The tunneling-specific interface resistivity can be calculated based on the Simmons tunneling diode model at low bias voltage \cite{simmons},
\begin{equation}
    \rho_{t} \approx \frac{2h^2 w_{\text{TB}}}{3e^2 \sqrt{2m\Phi_{\text{TB}}}} \mathcal{T}^{-1}.
\end{equation}
Although $\rho_t$ is a quantity detrimental to charge injection efficiency for metal/semiconductor contact application, its presence in dielectric/semiconductor interface offers a mechanism unique to vdWHs for suppressing the leakage current. 
For LaO$X$/$M$S$_2$ vdWHs, $\rho_t$ are found to be ranging from $10^{-9}$ to $10^{-10}$ $\Omega$cm$^{2}$, which is comparable to other metal/semiconductor contacts \cite{nat_bi, wang_npj} (see Table IV). 

\subsection{Leakage Current: Interfacial Charge Injection and Interface Potential Barrier Tunneling}

To assess the capability of LaO$X$ as gate dielectrics to 2D TMDC, we calculate the leakage current of LaO$X$/$M$S$_2$ using the DFT-calculated band offset values. The leakage current is dominantly contributed by transport mechanisms \cite{infomat, PRL_ang}: (i) semiclassical overbarrier thermionic emission (TE); and (ii) quantum mechanical field-induced emission (FE). Based on the Murphy-Good model \cite{MG}, the leakage current density is
\begin{equation}
\begin{aligned}
        \mathcal{J}_{ \text{MG} } = {}  &\ \frac{4\pi em k_B T}{h^3} \int_{-W_a}^{W_l}  \frac{\ln\left[1+\exp\left(\frac{- E_{\text{k}}^{\text{(DFT)}} -W}{k_B T}\right)\right]}{1+\exp{\left[\frac{4\sqrt{2m^* W^3}}{3e\hbar F}v(y)\right]}} \,dW \ \\
        & + \frac{4\pi em k_B T}{h^3} \int_{W_l}^{\infty} \ln\left[1+\exp\left(\frac{-E_{\text{k}}^{\text{(DFT)}} - W}{k_B T}\right)\right]\,dW \ ,  \ 
\end{aligned}
\end{equation}
where the subscript $k = (\text{CBO}, \text{VBO})$ for leakage current flowing in the conduction band and the valance band, respectively, $k_B$ is the Boltzmann constant, $T$ is the temperature, $F$ is the applied field, $-W_a$ is the effective constant potential of electrons in the metal contact and can be taken as $ -W_a \to -\infty$ for most metals, $W_l = -\sqrt{e^3F/8\pi\epsilon_{\perp, 0}}$, $y = \sqrt{e^3F/4\pi\epsilon_{\perp, 0}}/|W|$, $v(y)$ is an elliptical function arising from image-charge effect \cite{MG} and $\epsilon_{\perp, 0}$ is the out-of-plane static dielectric constant of the dielectric layer. 
The first and second term in the right-hand side of Eq. (6) corresponds to the underbarrier quantum mechanical field-induced tunneling and the overbarrier semiclassical thermionic emission, respectively.
We assume a planar diode geometry, i.e. $F = V/d_{t}$, where the tunneling distance is $d_{t}$ and the potential difference across the dielectric is $V = 0.355$ V based on the IRDS specifications of supply power voltage $V_{\text{dd}} = 0.70$ and saturation voltage $V_{\text{sat}} = 0.345$ V \cite{NC2021,IRDS}. 
Here $m$ is the free-electron mass and $m^*$ is the effective electron or hole masses of LaO$X$ for leakage current conduction mediated by conduction band or by valance band, respectively. 
Based on previous DFT calculations \cite{NC2021}, the electron and hole masses of LaOCl (LaOBr) is 1.125 $m$ (2.251 $m$) and 1.475 $m$ (1.180 $m$), respectively. 
The thickness of LaOCl and LaOBr monolayer is $t_{\text{LaOCl}} = 0.708$ \AA and $t_{\text{LaOBr}} = 0.794$ \AA, respectively \cite{NC2021}.
The tunneling distance is then estimated as $d_t = t_{\text{LaO}X} + 2 d_{\text{vdW}}$ where we assume that the LaO$X$ monolayer forms van der Waals contacts at both of its top and bottom surfaces with a typical van der Waals gap distance of $d_{\text{vdW}} = 3$ \AA. 

Equation (6) can be simplified into a `tunneling' and a `thermionic' component to yield a simplified leakage current formula commonly used in the literature \cite{NC2021, 2D_vdw_dielectric2}, i.e. $\mathcal{J}_0 = \mathcal{J}_{\text{tun}}+\mathcal{J}_{\text{therm}}$, where
\begin{subequations}
    \begin{equation}
        \mathcal{J}_{\text{tun}} = \frac{e^3}{8\pi h E_{\text{k}}^{\text{(DFT)}}}{F}^2\exp\left(-\frac{4\sqrt{2m^*}{E_{\text{k}}^{\text{(DFT)}3/2}}}{3e\hbar{F}}\right),
\end{equation}
\begin{equation}
        \mathcal{J}_{\text{therm}} = \frac{4\pi em {k_B}^2 }{h^3}T^2\exp\left(\frac{-\left(E_{\text{k}}^{\text{(DFT)}3/2}-\sqrt{{e^3F}/{4\pi \epsilon_{\perp,0}}}\right)}{k_B T}\right).
\end{equation}
\end{subequations}
In Table IV, we calculate the leakage current using both the Murphy-Good and the simplified models. In general, the simplified model \emph{underestimates} the leakage current.
The underestimation can reach about 3 orders of magnitude (i.e. CBO leakage currents of LaOCl/WS$_2$ and LaOBr/WS$_2$, and VBO leakage currents of LaOCl/MoS$_2$ and LaOCl/WS$_2$), thus highlighting the importance of using Murphy-Good model to obtain a more accurate estimation of the leakage currents. 
The CBO leakage currents of LaOCl/WS$_2$ and LaOBr/WS$_2$ are higher than the $10^{-2}$ Acm$^{-2}$ threshold proposed in \cite{2D_rev2}, thus suggesting that LaO$X$ may not be suitable for NMOS configurations in WS$_2$-based devices. 
Similarly, the sizable VBO leakage current of LaOBr/MoS$_2$ reveals the incompatibility for PMOS application. 
Nevertheless, the CBO and VBO leakge currents of LaOCl/MoS$_2$ are both in the orders of 10$^{-7}$ and $10^{-8}$ Acm$^{-2}$, respectively, which are significantly lower than the $10^{-2}$ Acm$^{-2}$ threshold leakage current threshold. 
LaOCl/MoS$_2$ is thus a versatile dielectric/semiconductor combinations that are simultaneously compatible for both NMOS and PMOS operations, while LaO$X$/WS$_2$ combinations are generally compatible with PMOS operations. 
If we relax the leakage current limit from $10^{-2}$ Acm$^{-2}$ \cite{2D_rev2} to the IRDS 2020 standard \cite{IRDS} of 0.145 Acm$^{-2}$ for a device of 28 nm pitch, 18 nm long gate, and an effective gate of 107 nm as proposed in \cite{2D_vdw_dielectric2}, then both LaOCl/MoS$_2$ and LaOBr/MoS$_2$ are simultaneously compatible with NMOS and PMOS device configurations. 

\section{Conclusion}

In summary, we performed DFT calcualtions to investigate the interfacial properties of dielectric/semiconductor van der Waals heterostructures composed of LaO$X$ ($X=$ Cl, Br) and $M$S$_2$ ($M=$Mo, W) monolayers. 
We found that the contact heterostructures exhibit negligible hybridization and interfacial charge transfer, thus completely preserving the electronic properties of the contacted $M$S$_2$ TMDC monolayers.
Importantly, the conduction and valance bands edges exhibit sizable band offsets in LaOCl/MoS$_2$ contact, thus suggesting its compatibility for both PMOS and NMOS device configurations, while LaO$X$/Br are compatible with PMOS device configuration. 
Together with the previously predicted exceptional out-of-plane static dielectric constants of LaO$X$ monolayers, our results suggest that LaOCl and LaOBr are excellent gate dielectric layered materials for MoS$_2$ and WS$_2$. 
This work provides theoretical foundation and interface physics insights that are useful for the design of novel 2D semiconductor devices utilizing atomically thin LaO$X$ as gate dielectrics, and shall provide a practical motivation for the experimental searches of LaOCl and LaOBr monolayers. 

\section*{Acknowledgement}
This work is supported by Singapore Ministry of Education (MOE) Academic Research Fund (AcRF) Tier 2 (MOE-T2EP50221-0019). Z.J. and L.K. Ang are supported by ASTAR AME IRG (A2083c0057)

\end{document}